\documentclass[twocolumn,pra,showpacs,floatfix, preprintnumbers]{revtex4}
\usepackage{dcolumn}
\usepackage{graphicx}
\usepackage{epsf}
\usepackage{amsmath}
\usepackage{bm}

\begin{document}

\newcommand{\sts}{$1S$\,--\,$2S$}

\bibliographystyle{myprsty}

\title{Photoionization Broadening
of the \sts\ Transition in a Beam of Atomic Hydrogen}

\author{N.~Kolachevsky$^{1,2}$}
\author{M.~Haas$^3$}
\author{U.~D.~Jentschura$^3$}
\author{M.~Herrmann$^1$}
\author{P.~Fendel$^1$}
\author{M.~Fischer$^4$}
\author{R.~Holzwarth$^{1,4}$}
\author{Th.~Udem$^1$}
\author{C.~H.~Keitel$^3$}
\author{T.~W.~H\"{a}nsch$^{1,5}$}

\affiliation{**$^{1}$\hbox{Max--Planck--Institut f\"{u}r
Quantenoptik, Hans--Kopfermann--Str. 1, 85748 Garching,
Germany}\\}

\affiliation{**$^{2}$\hbox{P.N. Lebedev Physics Institute,
Leninsky prosp. 53, 119991 Moscow, Russia}\\}

 \affiliation{**$^{3}$\hbox{Max--Planck--Institut f\"{u}r Kernphysik, Saupfercheckweg 1, 69117 Heidelberg, Germany}\\}

\affiliation{**$^{4}$\hbox{MenloSystems GmbH, Am Klopferspitz 19,
82152 Martinsried, Germany}\\}

\affiliation{**$^{5}$\hbox{Ludwig--Maximilians--University,
Munich, Germany}\\}

\begin{abstract}

We consider the excitation dynamics of the two-photon \sts\
transition in a beam of atomic hydrogen by 243 nm laser radiation.
Specifically, we study the impact of ionization damping on the
transition line shape, caused by the possibility of ionization of
the $2S$ level by the same laser field. Using a Monte-Carlo
simulation, we calculate the line shape of the \sts\ transition
for the experimental geometry used in the two latest absolute
frequency measurements (M. Niering {\it et~al.}, PRL 84, 5496
(2000) and M. Fischer {\it et~al.},
 PRL 92, 230802 (2004)).
The calculated line shift and line width are in excellent
agreement with the experimentally observed values. From this
comparison we can verify the  values of the dynamic Stark shift
coefficient for the \sts\ transition for the first time on a level
of 15\%. We show that the ionization modifies the velocity
distribution of the metastable atoms, the line shape of the \sts\
transition, and has an influence on the derivation of its absolute
frequency.
\end{abstract}


\maketitle

\section{Introduction}

High-precision spectroscopy of the \sts\ transition in atomic
hydrogen and deuterium provides an essential contribution to the
determination of the Rydberg constant, the Lamb shift
\cite{Biraben}, the $2S$ hyperfine interval \cite{KolPRA} as well
as it allows to study nuclear properties of the proton
\cite{Weitz2s4s} and the deuteron \cite{Huberisotop}. Recently it
has been shown, that monitoring the absolute frequencies of narrow
clock transitions (like the \sts\ transition in hydrogen and
quadrupole transitions in the Hg$^+$ and Yb$^+$ ions) over a
prolonged time interval, opens an opportunity to set a stringent
restriction on the possible slow variation of the fine structure
constant \cite{Fischetal, Peik04}, testing very fundamental
aspects of physics.
 This laboratory approach relies on the accuracy of the
\sts\ frequency measurement, which has now reached the level of
$1.4\times10^{-14}$, but in principle has still room for
improvement, since the Q-factor of this transition is about
$2\times10^{15}$. The accuracy which is achievable experimentally
is restricted by a number of systematic effects which shift the
frequency and change the line shape. So a further increase of
accuracy requires not only an upgrade of the experimental setup
and an improvement of the signal-to-noise ratio, but also a
tighter control of the systematic effects. Among these, the
dynamic Stark shift, the second order Doppler effect and the
time-of-flight (TOF) broadening have been considered as the most
important issues until recently.

In this article, we analyze how the presence of ionization losses
of the $2S$ state via the absorption of 243 nm photons contributes
to the line shape of the \sts\ transition in the experimental
configuration used for the most recent absolute frequency
measurement \cite{Fischetal, NiEtAl2000}. Comparing the
experimentally obtained intensity-dependent line width and
frequency shift of the \sts\ transition with results of a
numerical simulation, we can also indirectly validate a number of
atomic parameters like the two-photon transition matrix element,
the AC Stark shift coefficient and the ionization coefficient,
which were recently re-derived in a unified formalism
\cite{Haasetal}. In addition, Monte Carlo simulations allow to
study systematic effects which are difficult to quantify
experimentally. From this we gain new insights on the origin of
the scatter in the \sts\ absolute frequency data observed in the
experiments of 1999 and 2003
 \cite{NiEtAl2000,Fischetal}.

The article is organized as follows: In Sec.~II, we
describe in brief the experimental apparatus emphasizing points
important for the current analysis, while the numerical model is
given in Sec.~III. We analyze the results in Sec.~IV,
followed by the conclusions.

\section{Experimental Setup}

The basic experimental setup used for the spectroscopy of the \sts\
transition in a cold beam of atomic hydrogen is described in detail
in \cite{HuGrWeHa1999}. In this section we will focus on a description
of the apparatus depicting the main features necessary for the current
analysis and will present the relevant experimental results.

The dipole-forbidden \sts\ transition is excited by means of a
Doppler-free two-photon driving scheme in counter-propagating
waves of equal frequencies.
 The necessary 243 nm radiation is produced by
doubling the cw light emitted by a 486 nm dye laser in a
$\beta$-barium borate crystal. The laser is stabilized to an
external ultra-stable cavity and has a spectral line width of
about 60 Hz while the drift of the laser frequency is less than
0.5 Hz/s \cite{Fischetal}.  The frequency of the 486 nm laser is
measured using the frequency comb technique \cite{Fischetal, ACFC}
which is described in detail in \cite{Comb}.

\begin{figure}[t!]
\begin{center}
\includegraphics[width=0.45\textwidth]{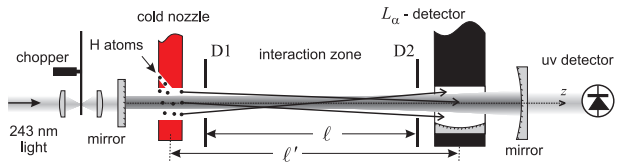}
\end{center}
\caption{Excitation of cold hydrogen atoms by 243 nm light in the
enhancement cavity. Atoms escaping the cold nozzle are collimated
by two diaphragms D1 and D2 before entering the detection zone.
All parts except the chopper assembly and the uv photodiode are
placed in vacuum of $10^{-7}$--$10^{-6}$ mbar.  The diameters of
the nozzle and the diaphragms are drawn not to scale and are much
smaller as suggested by this graph. } \label{fig1}
\end{figure}

To efficiently excite the \sts\ transition in atomic hydrogen we
enhance the 243 nm radiation in a linear cavity consisting of a
plane incoupling mirror (transmission $T_1=2.4 \times 10^{-2}$)
and a concave outcoupling mirror (radius $R=-4$ m, $T_2=1.36\times
10^{-4}$) separated by a distance of 29 cm as depicted in
Fig.~\ref{fig1}. The cavity loss is dominated by the diaphragms D1
and D2 that partially touch the laser mode so that the input
coupling mirror roughly corresponds to impendence matching
condition.  The radiation leaking out of the cavity is monitored
by a silicon photodiode which is periodically calibrated by a
power meter (Fieldmaster, Coherent Inc.). The photodiode readout
has been averaged over one second which is the laser dwelling time
when recording the spectra. Dividing the photodiode readout by the
transmission of the outcoupler $T_2$ we get the total power
circulating in the enhancement cavity per direction. The averaged
power is  recorded for each measured spectrum.

 Atoms produced in the microwave gas discharge  escape
a cold nozzle ($T\simeq5$ K) of 1.2 mm in diameter and enter the
interaction zone of length $\ell'=15.0$ cm, restricted by two
diaphragms D1 and D2 of respective diameters $2r_{\rm D1}=1.3$ mm
and $2r_{\rm D2}=1.4$ mm separated by $\ell=13.6$ cm.
During the flight through the Gaussian profile of the cavity
TEM$_{00}$ mode, a part of the atomic hydrogen is excited to the
metastable $2S$ state. Excited atoms are detected by quenching in
a dc electric field through the release of a Lyman-alpha photon
\cite{HuGrWeHa1999}.


To set an upper limit for the velocity of the atoms that
contribute to the signal and thus to reduce the influence of the
velocity-dependent systematic effects, a time-of-flight detection
technique \cite{HuGrWeHa1999, Fischetal} is used.
All atoms with velocities higher than $v_{\rm max}$ escape the
detection zone before the start of the detection and the
contribution of these atoms is excluded from the resulting
time-delayed spectrum. The cut-off velocity is defined as $v_{\rm
max}=\ell'/\Delta\tau$, where $\Delta\tau$ is a delay between
closing of the 243 nm laser light and the start of detection.


The 486 nm laser is scanned stepwise across the resonance and for
each laser frequency a multichannel scaler records the number of
detected Lyman-$\alpha$ photons for 12 different delays
$\Delta\tau_i=10,\ 210, \ 410,\ \ldots,\ 2210\ \mu s $ in the
corresponding time windows \{$\Delta\tau$, 3 ms\}.

Using this detection technique we simultaneously record 12 lines,
each containing the contributions of atoms in the respective
velocity range $0<v<v_{\rm max}(\Delta\tau_i)$ ($i=1,\ldots,12$),
selected from the same initial velocity distribution $f(v)$. As
follows from \cite{Maxwdistrib}, the velocity distribution in a
one-dimensional thermal atomic beam effusing from thermalized gas
volume through a hole in a thin wall is given by:
\begin{equation}\label{maxw}
f(v)\propto (v/v_0)^3\exp[-(v/v_0)^2]\,,
\end{equation}
where $v_0=\sqrt{2k_BT/m_{\rm H}}$ is the most probable thermal
velocity for a given temperature $T$ and $m_{\rm H}$ is the mass
of hydrogen atom. As discussed in Ref.\,\cite{Maxwdistrib} and
references therein, expression (\ref{maxw}) is valid for perfectly
collimated beams and should be further modified in other cases
also because the thin wall assumption is not fulfilled in our
case. The collimation angle of the atomic beam in our experiment
is about $0.01$ which justifies the use of the mentioned
approximation.The impact of this approximation on the \sts\
frequency shift will be discussed in Section \ref{chfreq}.

The increase of the delay $\Delta\tau$ reduces the maximum
velocity $v_{\rm max}(\Delta\tau)$ of atoms contributing to the
signal and decreases the Lyman-$\alpha$ count rate.
Correspondingly, the contributions of the time-of-flight
broadening and the second-order Doppler effect are reduced, which
in turn leads to a narrowing and a symmetrization of the \sts\
transition line shape. Hence, at the cost of a lower
signal-to-noise ratio, lines with lower systematics are recorded.
Using this experimental approach and the line-shape model
\cite{HuGrWeHa1999} one can correct for the second order Doppler
effect which is necessary for the accurate determination of the
absolute \sts\ transition frequency.

In this paper we will study three types of intensity related line
broadening effects: Because the  $2S$ state is close enough to the
continuum that a single additional 243 nm photon is sufficient to
ionize the hydrogen atom reducing the effective interaction time.
In addition the spatially dependent AC Stark shift and the usual
power broadening caused by Rabi flopping are present.

For simplicity, we will analyze only hydrogen lines recorded at a
particular delay, $\Delta\tau=1210 \ \mu$s which corresponds to
$v_{\rm max}\simeq120$ m/s: this is a compromise between simple
line shape and sufficient signal-to-noise ratio. At this delay the
line shape can be well approximated by a Lorentzian and the
maximum second-order Doppler effect does not exceed $200$ Hz in
modulus, the typical line width being 1-2 kHz at 121 nm. Therefore
we consider contributions to the width of a Lorentzian line, which
provides a much clearer insight into the nature of the different
broadenings than the complete line shape model for all $\Delta
\tau$.

\begin{figure}[t!]
\begin{minipage}[b]{0.49\textwidth}
\begin{center}
\includegraphics[width=0.85\textwidth]{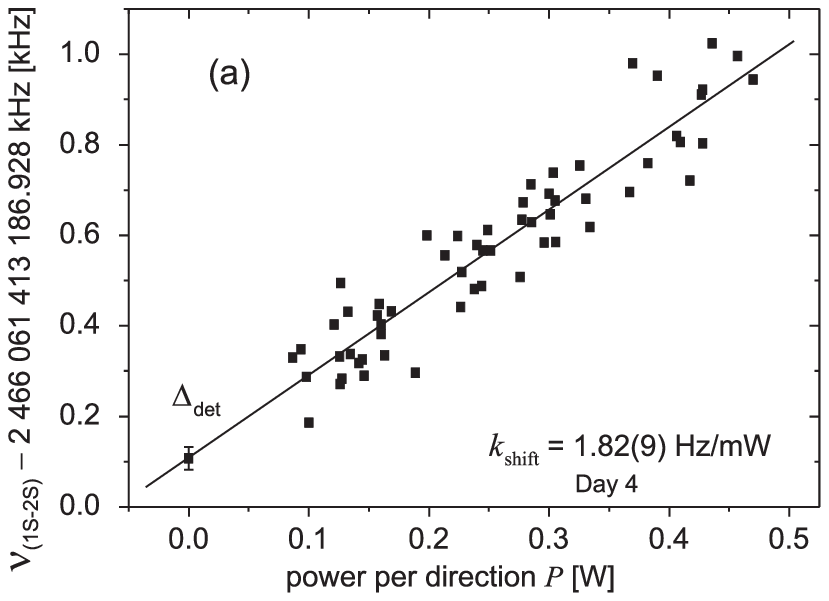}
\end{center}
\end{minipage}

\vspace{0.6cm}
\begin{minipage}[b]{0.49\textwidth}
\begin{center}
\includegraphics[width=0.85\textwidth]{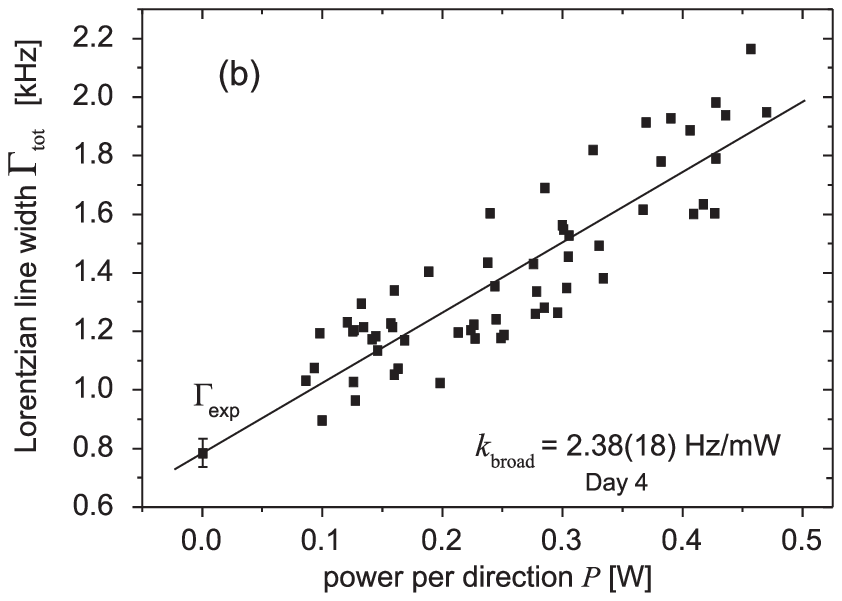}
\end{center}
\end{minipage}
\caption{Results from 59 line scans: (a) center frequencies and
(b) line widths (FWHM) both derived from Lorentzian fits versus
the excitation power circulating in the cavity per direction, $P$.
The power of 1 W per direction corresponds to a total on-axis
focal intensity of 16~MW/m$^2$ averaged over one wavelength of the
standing wave.
} \label{fig2}
\end{figure}

To correct for another important systematic effect, the dynamic Stark
shift, we recorded a number of \sts\ spectra at different excitation
powers during each day of measurement \cite{Fischetal}. Here we use the
same set of data for the analysis.

All lines recorded at $\Delta\tau=1210 \ \mu$s have been fitted by
Lorentzian functions and then the absolute frequencies of the line
centers $\nu_{\rm{1S-2S}}$ and line widths $\Gamma_{\rm tot}$ were
plotted versus the excitation light power $P$ (per direction). The
results for the data set recorded on one of the days of
measurement and the linear extrapolation to zero intensity are
presented in Fig.~\ref{fig2} as an example.

The intensity-dependent line shift $\nu_{\rm{1S-2S}}(P)$ is mainly
caused by the dynamic Stark shift \cite{Haasetal} with small
contributions of velocity-selective ionization. As in Refs.
\cite{Fischetal, NiEtAl2000} we fit the data with a linear
function (since for low intensities the AC Stark shift for \sts\
transition is proportional to the intensity), but instead of
extracting just the absolute frequency at zero power we also
determine the slope of the fit, $k_{\rm{shift}}$, as shown in the
figure. Correspondingly, we consider the slope of the linear fit
of the $\Gamma_{\rm tot}(P)$ dependence and call it
$k_{\rm{broad}}$. It will be shown in the following that this
coefficient is formed by three important contributions, namely
power broadening, inhomogeneous AC Stark shift in a Gaussian beam
profile and the influence of the ionization of the $2S$ level.

In the next section we will analyze the structure of these coefficients
by the help of a numerical analysis performed for an ensemble of hydrogen
atoms excited in the given experimental geometry.

\section{Excitation of a beam of atomic hydrogen in the presence of
ionization}

The calculation of the $2S$ excitation dynamics is based on the
equations of motion for the matrix elements of the density
operator $\rho'$ in the rotating wave approximation
\cite{Haasetal}. The laser field of angular frequency
\mbox{$\omega_L\simeq 2\pi\times\,1233$ THz}, corresponding to
$\lambda=243$~nm, and intensity $I$ excites the two-photon
transition between the ground state $|g\rangle$ ($1S$) and the
excited state $|e\rangle$ ($2S$). The frequency detuning is
defined to be the difference between the doubled laser frequency
and  the resonance frequency
$\Delta\omega=2\omega_L-\omega'_{eg}$, where $\hbar\omega'_{eg}$
is the energy difference between the $1S$ and $2S$ states of a
perturbed atom (see Eq.~(\ref{detuning})), ignoring the  the
hyperfine structure~\cite{KolPRA}.

\begin{figure*} [t!]
\begin{minipage}[b]{0.33\textwidth}
\begin{center}
\includegraphics[width=0.95\textwidth]{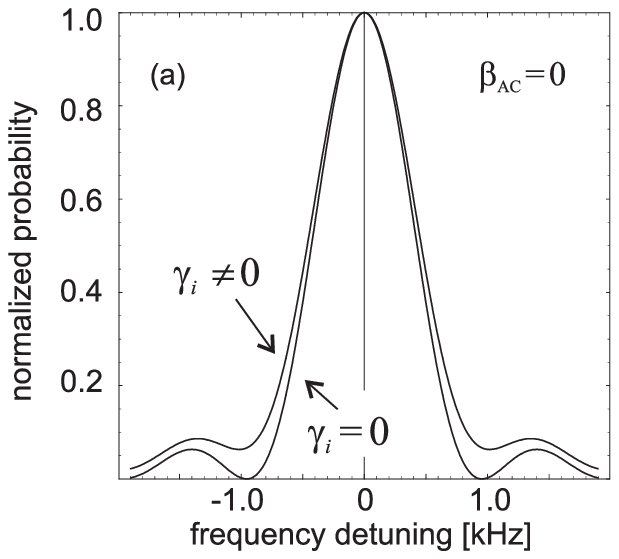}
\end{center}
\end{minipage}
\begin{minipage}[b]{0.66\textwidth}
\begin{center}
\includegraphics[width=0.95\textwidth]{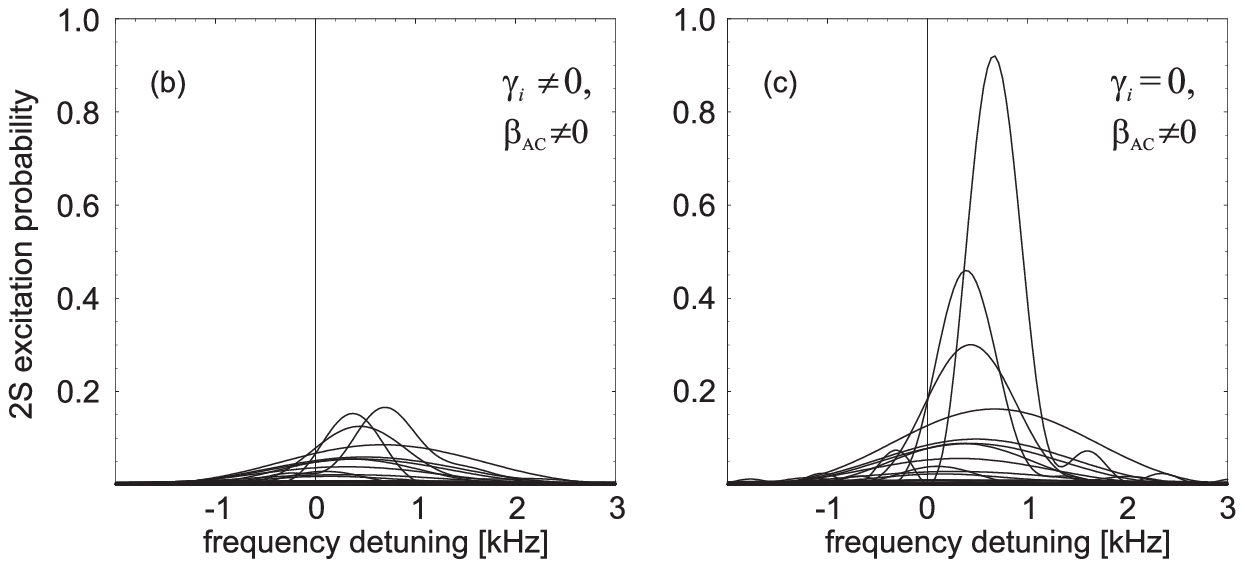}
\end{center}
\end{minipage}
\caption{(a) -- normalized  responses of a single atom illuminated
by a 1 ms rectangular pulse with the intensity of $I=4$ MW/m$^2$
with ionization channel ($\gamma_{\rm i}\neq0$) and without it
($\gamma_{\rm i}=0$).
 For
simplicity, neither the AC Stark shift nor the Doppler effect are
taken into account. (b), (c) -- excitation probabilities for atoms
flying along different random trajectories through the Gaussian
intensity profile corresponding to $P=300$ mW. (b) -- calculations
for 50 randomly chosen  trajectories in presence of the ionization
channel. (c) -- calculation for the same ensemble of trajectories
without ionization taken into account.
  } \label{figmax}
\end{figure*}

In the experiment, the excitation field intensity $I$ depends on
the position inside the excitation region, because the laser
profile is of Gaussian type. The fact that each atom flies along a
certain trajectory with an individual speed through this intensity
profile, renders the intensity a time dependent quantity $I(t)$ in
the frame of the travelling atom. Because the dynamic Stark effect
is proportional to the intensity, also the total detuning as
measured by the atom, is a function of time, $\Delta\omega(t)$. In
addition the upper state $|e\rangle$ decays to the continuum due
to photoionization with a time-dependent rate proportional to the
laser intensity.
One can write the equations of motion for the density
matrix, neglecting spontaneous decay of the $2S$ state
\cite{Haasetal}. After factoring out the fast optical oscillations
from the components of the density matrix $\rho$ via $\rho'=\rho
\exp(-2i\omega_Lt)$ we obtain the two-photon Bloch equations:
\begin{subequations}
\begin{eqnarray}
\frac{\partial}{\partial t}\rho'_{gg}(t)&=&-\Omega(t)\,
{\rm{Im}}(\rho'_{ge}(t)),\\\nonumber \rule{0pt}{5ex}
\frac{\partial}{\partial
t}\rho'_{ge}(t)&=&-{\rm{i}}\,\Delta\omega(t){\rho}'_{ge}(t)+
{\rm{i}}\frac{\Omega(t)}{2}({\rho}'_{gg}(t)-{\rho}'_{ee}(t))\\
&&-\frac{\gamma_{\rm i}(t)}{2}{\rho}'_{ge}(t),\\
\rule{0pt}{4ex}
\frac{\partial}{\partial t}\rho'_{ee}(t)&=&\Omega(t)
{\rm{Im}}(\rho'_{ge}(t))-\gamma_{\rm i}(t)\rho'_{ee}(t),
\end{eqnarray}
\label{EOMdensityMatrix}
\end{subequations}
where the time-dependent two-photon Rabi frequency is
$\Omega(t)=2(2\pi\beta_{ge})(m_e/\mu)^3I(t) $. Here $m_e$ is the
electron mass, $\mu$ the reduced mass, and the value of
$\beta_{ge}$ is taken from \cite{Haasetal}:
$\beta_{ge}=3.68111\times10^{-5}$~Hz(W/m$^2$)$^{-1}$ for the \sts\
transition. In the same paper the ionization rate is given by
$\gamma_{\rm i}(t)=2\pi\beta_{\rm ioni}(e)(m_e/\mu)^3 I(t)$, where
the ionization coefficient reads $\beta_{\rm ioni}(e)
=1.20208\times10^{-4}$ Hz\,(W/m$^2$)$^{-1}$.

Experimentally an ensemble of atoms contributes to the signal,
with each individual atom having a different speed and trajectory
through the interaction zone. We take into account only those
trajectories, which are confined by the diaphragms D1 and D2 (see
Fig.~\ref{fig1}).  It is assumed, that the starting points on the
diaphragm D1 and the exit points on the diaphragm  D2 defining
each trajectory are randomly and independently distributed over
the corresponding diaphragm with constant area density. It means,
that the angle $\varphi>0$ between the cavity axis $z$ and each
individual trajectory is also randomly and uniformly distributed
for $r_{\textrm{D1}},\ r_{\textrm{D2}}\ll\ell$: the atom can
intersect the excitation light beam as well as fly parallel to its
axis. The velocities are also chosen at random according to a
Maxwellian distribution like in Eq.~(\ref{maxw}) at $T=5$ K, as in
the experiment the day-averaged readouts of the temperature
sensor, placed on the cold nozzle, ranged between 4.95 K and 5.25
K.

After this longitudinal velocity seeding, all atoms which would
reach the detector earlier than the time at which the delayed
detection started (that is before $\Delta\tau=1210\ \mu$s), were
rejected from the ensemble. For each remaining atom, one can then
easily calculate the time-dependent intensity $I(t)$, as well as
the total interaction time. The laser is modelled by a Gaussian
beam profile with parameters defined by the experimental geometry.
The radial intensity distribution is therefore given by
$I(r)=(4P/\pi w^2)\,{\rm{exp}}[-2r^2/w^2]$, where
$w^2=w_0^2(1+(z\lambda/\pi w_0^2)^2)$.
 The beam slightly
diverges along the $z$-axis, and the beam waist on the incoupler
plane equals $w_0=283\ \mu$m. Here $P$ is  the power per
direction, as used in Fig.~\ref{fig2}.

It is necessary to point out, that atoms flying through a standing
wave in the cavity interact with a highly inhomogeneous laser
field consisting of nodes and maxima with the periodicity of 121.6
nm in $z$-direction. But even for a very slow atom travelling at a
longitudinal velocity of $v=v_0/300\simeq1\ \rm{m/s} $,
Eq.~(\ref{maxw}), the intensity will be modulated at a frequency
of about 10 GHz, which is 6 orders of magnitude higher than the
typical Rabi frequency $\Omega$, which governs the population
dynamics. Thus for the calculation of the density matrix elements,
we can equivalently use a laser field with an averaged intensity,
in a similar way as in applying the rotating wave approximation.

The general detuning $\Delta\omega(t)$ along the trajectory can now be
expressed as
\begin{eqnarray}\label{detuning}
  \Delta\omega(t) &= &2\omega_L - \omega'_{eg}=\\ \nonumber
   &=&2\omega_L-\omega_{eg}+
  \frac{\omega_{ge}}{2}\left(\frac{v}{c}\right)^2
  -2\pi\,\beta_{\rm{ac}}\,I(t),
\end{eqnarray}
where $\hbar\omega_{eg}$ is the energy between the unperturbed
$1S$ and $2S$ states. The other summands take into account  the
second order Doppler effect and the intensity-dependent frequency
shift caused by the real part of the dynamic Stark effect. In
turn, the imaginary part of the dynamic Stark effect is related to
ionization. According to \cite{Haasetal}, $\beta_{\rm{ac}} =
[\beta_{\rm ac}(2S)-\beta_{\rm ac}(1S)](m_e/\mu)^3 =
1.66982\times10^{-4}$ Hz\,(W/m$^2$)$^{-1}$.

The time evolution of Eqs.~(\ref{EOMdensityMatrix}) for a specific
atom and its corresponding interaction time is calculated
numerically for each given laser frequency $\omega_L$ which is
scanned stepwise across the resonance.
The resulting value of $\rho'_{ee}$ is stored
together with $\omega_L$ to define the contribution of each
individual atom to the line profile.

Fig.~\ref{figmax}a illustrates the spectral response of a single
atom at rest to the excitation with the rectangular pulse of 243
nm radiation. This case is analyzed in detail in \cite{Haasetal}.
One observes, that for moderate excitation intensities of a few
MW/m$^2$ and for short interaction times less then 1 ms, the
contribution of the ionization broadening to the total line width
is practically negligible. Indeed, for a 1 ms rectangular pulse of
intensity  $I=4$ MW/m$^2$  the ionization broadening is a very
small effect, though the ionization rate itself in this case
equals $\gamma_{\rm i}=480$~Hz.

Fig.\,\ref{figmax}b illustrates how individual atoms flying
through the excitation region contribute to the observed line
shape.
Depending on the individual trajectory and the velocity of an atom
it will be exposed to a time varying intensity. The figure shows
the resulting individual line shapes, over which the detector
integrates to yield the observed line profile. Note the variations
in amplitude, shift and width of the individual lines due to
varying speeds (second order Doppler shift), interaction times
(excitation and ionization probability, TOF broadening) and
sampled intensities (excitation and ionization probability, AC
Stark shift, power broadening).

 In contrast to the physical case given in Fig.~\ref{figmax}b with all effects
taken into account, Fig.~\ref{figmax}c shows the excitation
probabilities for \emph{the same ensemble} of 50 trajectories
calculated without ionization. Apparently the ionization
preferably removes the slow atoms from the initial distribution,
because they have more time to interact with the field. But  these
atoms would have contributed the strongest and the narrowest
spectral lines to the resulting collective spectrum if no
ionization had taken place. As a result, the detected line becomes
weaker and broader.

 In the
results section we will discuss the simulated line profiles in more
detail and compare the results with experimental data.

To perform the described Monte-Carlo simulation we used the
package ``Mathematica 5'' \cite{Matematica}. The simulation has
shown good convergence and an arbitrary ensemble of 10\,000 atoms
has been used to calculate each transition line shape. The input
parameters used by the program can be divided into four groups:
(i) atomic structure, (ii) geometric parameters ($w_0$, $r_{\rm
D1}$, $r_{\rm D2}$, $\ell$, $\ell'$, nozzle radius), (iii)
experimental conditions ($P$, $T$, $\Delta\tau$) and (iv) two
program switches which allowed to artificially set $\gamma_{ i}=0$
(no ionization) or $\beta_{\rm ac}=0$ (no AC Stark shift).
 The last option allows to analyze separately the
contributions of ionization, AC stark shift and power broadening
to the line shape.

Including all contributions to $\rho'_{ee}(\Delta\omega_L)$, we
obtain the simulated line shape, which we fit with a Lorentzian
profile exactly like in the evaluation of the experimental data
for the purpose of this study. In our simulations, we varied the
excitation power up to $P=1.2$ W per direction which covers and
exceeds the whole experimental range.

\section{Results }

\begin{figure*}[t!]%
\begin{center}
\includegraphics[width=0.8\textwidth]{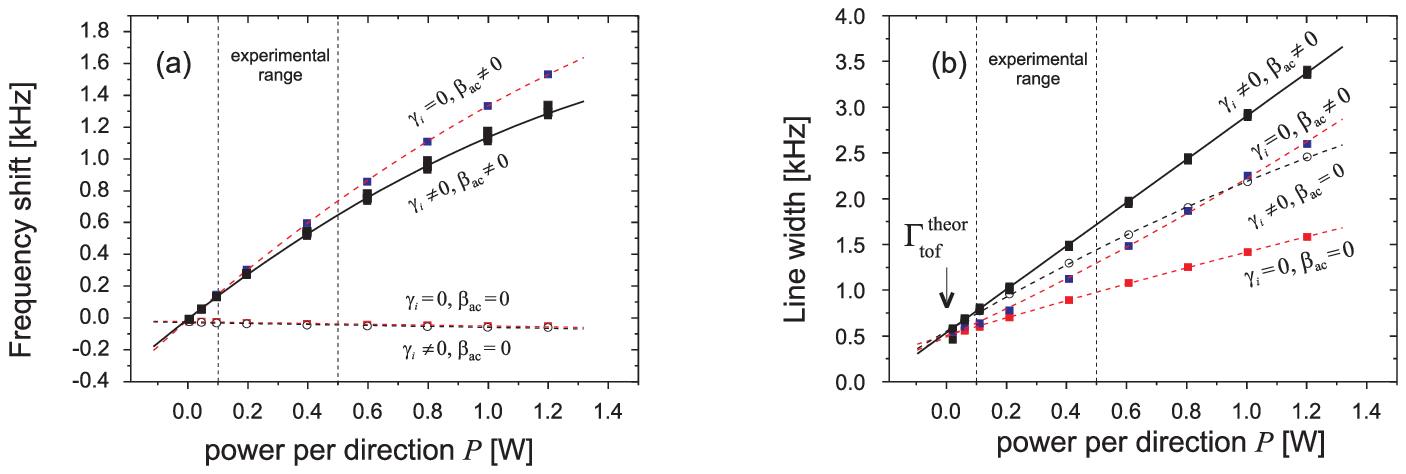}

\vspace{0.02\textheight}
\includegraphics[width=0.8\textwidth]{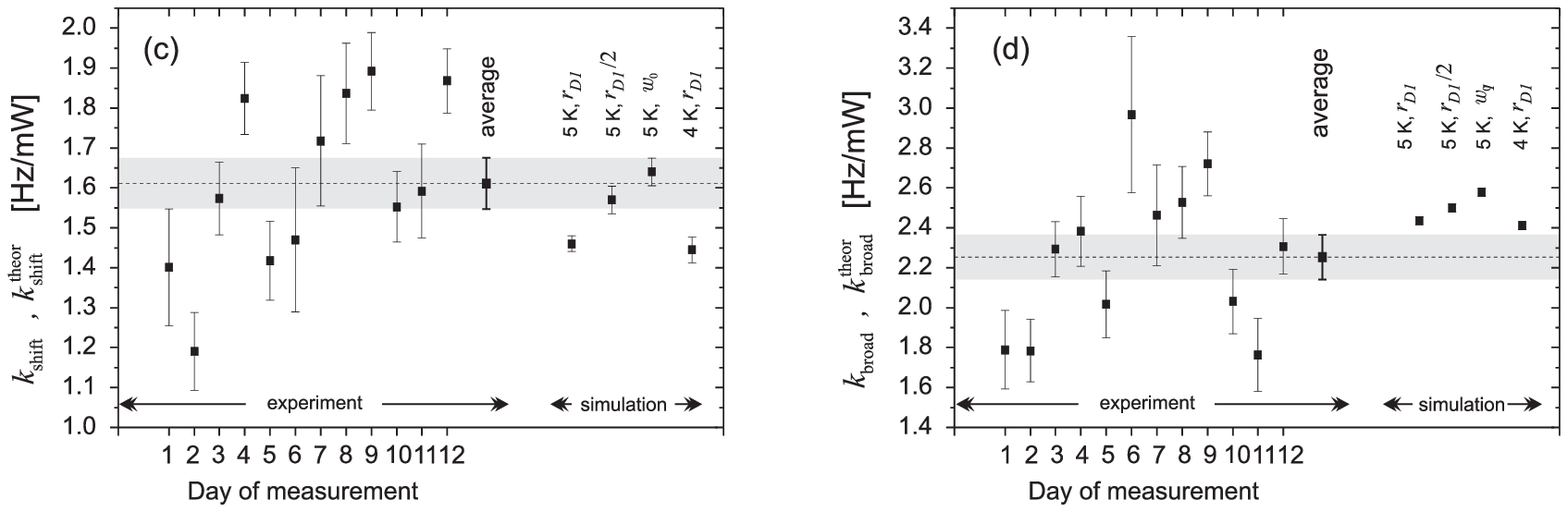}
\end{center}
\caption{Simulated dependencies of the frequency shift (a) and the
line width (b) of the \sts\ transition vs. excitation power per
direction, $P$ and a detection delayed by $\Delta \tau = 1210\
\mu$s. The set of data marked $\beta_{\rm ac}\neq0,\ \gamma_{\rm
i}\neq0$ corresponds to the physical case. The other simulations
clarify the separate contributions of ionization, inhomogeneous AC
Stark shift and power broadening (e.g. the data set with
$\gamma_\textrm{i}=0,\ \beta_{\textrm{AC}}=0$ in (b) corresponds
to pure  power broadening). Note that the contributions cannot
 be added  in any simple algebraic way. The data sets are
fitted by parabolic functions.
 On the left side of (c) and
(d) the  the day-averaged values of $k_{\rm shift}$  and $k_{\rm
broad}$  with corresponding uncertainties are presented. The total
12-day average of the data is indicated by the dashed lines and
the gray bars for the uncertainties.
Simulations for two
different beam temperatures (5 K and 4 K) and for different radii
of the entrance diaphragm D1 ($r_{\rm D1}$, $r_{\rm D1}/2$ and
$w_0$)
 are presented on the right part of the lower plots.
   } \label{fig3}
\end{figure*}

In this section the results of the numerical simulation will be
compared with experimental results obtained in the measurement
\cite{Fischetal}. We will analyze the frequency shift and the line
shape of the \sts\ transition  and  discuss possible origins of
the excess data scattering observed in the frequency measurement
experiments.

We would like to point out here, that both the frequency shift and
the line broadening are \emph{very small} effects of the order of
$10^{-12}$ compared to the optical carrier frequency. Only the
 improvements of the laser stability and the narrowing of its
spectral line width along with an improved absolute frequency
measurement have made it possible for the first time to perform a
\emph{quantitative} analysis of the line shift and especially of the
line broadening as a function of excitation intensity.

Figure~\ref{fig3} illustrates the results of the different
simulations (a, b) and their comparison with experimental
observations (c, d). To evaluate the uncertainty of the simulation
for 10\,000 atoms due to the random elements of the simulation, we
performed the simulation for 3 different arbitrary ensembles of
atoms for the same input program parameters. The variation is on
the level of few per cent which is sufficient for this analysis.

\subsection{Line Center Shift}

As follows from Fig.~\ref{fig3}a, the main origin of the frequency
shift is the AC Stark shift. The physical case is depicted in a
solid curve ($\gamma_{\rm i}\neq0,\ \beta_{\rm ac}\neq0$). The
line shift reveals an observable non-linearity at higher powers.
It could be explained by the following contributions: (i)  the
inhomogeneous AC Stark shift, (ii) the influence of ionization and
(iii) the saturation of the transition. In the current analysis we
took the non-linearity into account by fitting the simulations in
Fig.~\ref{fig3}a with a parabolic function.

For comparison with the full treatment, we performed the same set
of calculations for three non-physical cases: with the ionization
channel switched off $( \gamma_{\rm i}=0,\ \beta_{\rm ac}\neq0)$,
the AC Stark shift is switched off $( \gamma_{\rm i}\neq0,\
\beta_{\rm ac}=0)$, and both these effects are switched off $(
\gamma_{\rm i}=0,\ \beta_{\rm ac}=0)$. We see, that the main
contribution to the frequency shift is due to the real part of the
dynamic Stark shift. The ionization, which reduces the number of
atoms undergoing the highest perturbation in the intensive laser
field. It also influences the frequency shift, but only if the
dynamic Stark shift is present as well.  All data sets in
Fig.~\ref{fig3}a converge at zero power to the same value of
$-20(1)\;\mbox{Hz}$ which corresponds to the mean second order
Doppler shift for the given beam temperature of $T=5$ K and the
delay time $\Delta\tau=1210\ \mu$s. This value is considerably
smaller than the maximal shift of $-200$ Hz calculated for $v_{\rm
max}(\Delta\tau)$. This difference is caused by the fact that the
velocity distribution of atoms which contribute to the signal
$f'(v)$ is different from the velocity distribution of atoms in
the beam $f(v)$: neglecting additional losses it is given by
\begin{eqnarray}
f'(v)&\propto& f(v)(1-v/v_{\rm max}(\Delta\tau)),\  \ v< v_{\rm
max},\\ \nonumber
 f'(v)&=&0,\qquad \qquad \qquad \qquad \qquad  v\geq v_{\rm max}.
 \label{modmax}
\end{eqnarray}
Consequently, the Doppler correction must take this weighting into
 account. A further correction to the effective second
order Doppler effect is invoked by the preferred removal of slow
atoms. As seen in Fig.\ref{fig3}a (for $\gamma_i \neq0$,
$\beta_{\rm ac}=0$) this preference and thus the associated shift
is vastly independent of laser power for our experimental
conditions.
%
%
  As it is impossible in our experiment to
measure the Stark coefficients $\beta_{\rm ac}$ and $\beta_{\rm
ioni}$ separately, we analyze the  values which can be derived
both from the experiment and our numerical analysis, namely the
slopes $k_{\rm shift}$ and $k_{\rm broad}$ of the power
dependencies given in Fig.~\ref{fig2} and Figs.~\ref{fig3}a,\,b.

On the experimental side, we derive these values for each of the
12 days of measurement exactly in the same way as depicted in
Fig.~\ref{fig2}. The results for each day are presented in
Figs.~\ref{fig3}\,c,d. One observes a significant scatter of the
day-to-day data, with  $\chi_r^2\simeq3$ (the reduced $\chi^2$ or
Birge ratio \cite{Birge}), which could be partly ascribed to
errors on the power axis, coming from both the calibration of the
photodiode measuring the 243 nm power and from temporal variations
of the transmittance of the curved mirror in Fig.~\ref{fig1}.
Therefore we average these data without taking into account the
individual uncertainties and obtain $k_{\rm shift}=1.61(6)$ Hz/mW
and $k_{\rm broad}=2.25(11)$ Hz/mW.

Extracting the coefficient $k_{\rm shift}$  from the results of
the numerical analysis, depicted in Fig.~\ref{fig3}a, is not
straightforward, since the curve for the shift of the line center
are not linear and therefore $k_{\rm shift}$ is not strictly
defined. However, since the scatter in the experimental data is
larger than the calculated nonlinearities, it is enough on the
present level of accuracy to replace the parabolic fit by a linear
fit in the case of the frequency shift-intensity dependence
($k_{\rm shift}$), in order to evaluate the results in the same
way as in the experiment. To reduce the error introduced by this
procedure, we restrict the data points taken into account by the
fit to those which fall into the experimental range of powers:
$0.1\; {\rm W}<P< 0.5\;{\rm{W}}$. The value resulting from the
Monte Carlo analysis is denoted as $k^{\rm theor}_{\rm shift}$  in
analogy to the notation in Fig.~\ref{fig1}.

To test the sensitivity of the model to the experimental
conditions and to evaluate the corresponding uncertainties we also
performed a set of simulations for different beam temperatures,
varying between 4 K and 6 K, slightly different interaction
lengths and a reduced radius $r_1$ of the diaphragm D1. All these
tests show a low sensitivity  of results to these parameters which
allows to compare them with the  experimental data. We have
performed calculations of $k^{\rm theor}_{\rm shift}$ for
different realistic experimental parameters. The results of four
selected cases are given in Fig.~\ref{fig3}c: with an atomic beam
temperature of 4 and 5 Kelvin and for a diaphragm size matched and
centered to the laser beam to simulate the nozzle covered with a
molecular hydrogen film.

The last case corresponds to a possible situation, where a film of
molecular hydrogen symmetrically covers the nozzle from inside in
such a way that the  nozzle radius  equals either $r_{\rm D1}/2$
or $w_0=283\ \mu$m  (the film eventually touches the laser beam).
The frozen nozzle radially restricts the atomic trajectories on
the entrance to the interaction zone, which results in changing
the intensity profiles. The growth of the H$_2$ film was a
characteristic feature of the hydrogen spectrometer working at the
lowest beam temperature. The measurement cycle lasted about 15
min, after which the nozzle was heated to 20 K to melt the film.

Comparing the averaged value $k_{\rm shift}$ with the results of
the numerical simulation, we can establish an agreement on the
level of about 15\% uncertainty with respect to the absolute
values. As follows from the previous analysis, the
intensity-dependent frequency shift is mainly caused by the AC
Stark shift which contributes to the general detuning
(\ref{detuning}). Using the linearity of the effect at low powers
one can conclude, that this analysis allows to verify $\beta_{\rm
ac}$ at the same confidence level of 15\% and therefore we can
experimentally verify the results of the calculations of
$\beta_{\rm ac}=\beta_{\rm ac}(2S)-\beta_{\rm ac}(1S)$ presented
in \cite{Haasetal}.



\subsection{Line Width}

The analysis of the line width is  more complicated in comparison
with the analysis of the frequency shift since a number of
different effects contribute significantly to the line broadening.
The line width consists of intensity-independent and
intensity-dependent parts. The intensity-independent part includes
the laser spectral line width and the time-of-flight broadening,
while  such effects as power broadening, inhomogeneous AC Stark
broadening and ionization broadening contribute to the latter
part. In addition, temporal intensity fluctuations of the laser
radiation in the excitation region also induce an additional
intensity-dependent broadening due to the variable Stark shift.
Thus, assuming for simplicity the Lorentzian profiles, one can
write:

\begin{equation}
 \Gamma_{\rm tot}(P)=\Gamma(P)+\Gamma_{\rm tof}+4\,\Gamma_{\rm laser}\,,
  \label{LaserLinewidth}
\end{equation}
where $\Gamma_{\rm tot}(P)$ is the total line width, $\Gamma_{\rm
tof}$ is the contribution of the time-of-flight broadening, and
$\Gamma(P)$ includes all intensity-dependent effects with
$\Gamma(0)=0$. The laser line width $\Gamma_{\rm laser}$ carries a
factor of 4 because we measure it at 486~nm whereas all other
transition frequencies correspond to 121 nm. It has been shown
experimentally, that the spectral line width of our laser system
doubles for the frequency doubling process. We understand that
each of the contributions listed above has a complex spectral
shape which can significantly differ from Lorentzian and the
expression (\ref{LaserLinewidth}) can be used only for  rough
estimations. Still, the simulations show that in our case the
time-of-flight broadening has near-to-Lorentzian spectral line
shape while heterodyne experiments indicate the same for the
spectral line of our laser system.

As is evident from Fig.~\ref{fig3}b, the intensity-dependent part
of the line width is formed by   three different processes: (i)
the power broadening, (ii) the inhomogeneous AC Stark shift in the
laser beam profile and (iii) the ionization of the $2S$ state. The
pure power broadening corresponds to the case $( \gamma_{\rm
i}=0,\ \beta_{\rm ac}=0)$ and is due to Rabi flopping between the
ground and the excited states, effectively reducing the lifetime
of the population in the excited $2S$ state. The inhomogeneous AC
Stark shift in the case $( \gamma_{\rm i}=0,\ \beta_{\rm
ac}\neq0)$ originates from a Gaussian intensity profile in the
excitation region. The last significant broadening comes from
ionization which contributes in a highly nontrivial way
illustrated in Figs.~\ref{figmax}b,c.

The excessive loss of slow atoms has been  taken into account in
the line form model \cite{NiEtAl2000,HuGrWeHa1999} as an empirical
modification of the low-velocity wing of the Maxwellian
$v^3$-dependency in Eq.~(\ref{maxw}). There it was assumed that
the slow atoms are missing from the beam because of collisions
with the fast atoms (Zacharias effect). However, now we conclude
that no such mechanism is necessary to fit our data and the
selective ionization is the most significant effect responsible
for this loss. Strictly speaking, one deals with three distinct
velocity distributions: (i) the atoms emerging from the nozzle,
(ii) the detected $2S$ atoms, and (iii) the ground state atoms.
Whereas the first distribution is assumed to be Maxwellian
(Eq.~(\ref{maxw})), the second is relevant for determining the
line shape.

To compare with the experimental results, we also evaluated the
contribution of the temporal intensity fluctuations. In principle,
this effect is equivalent to the considered inhomogeneous AC Stark
shift, but integrated over time, not space. In the experiment, we
observed fast intensity fluctuations on the level of $\pm5\%$ during
the excitation cycle. We embedded this process into the simulation
by adding white noise with a $5\%$ band to $I(t)$.  The effect lies
within the uncertainty of the calculations for the given number of
atoms used in the simulation.

For the line width as a function of laser power, the slope is
derived  from the  fit similar to the case of Fig.\,\ref{fig3}a.
The value resulting from the Monte Carlo analysis is denoted as
$k^{\rm theor}_{\rm broad}$. The line width, extrapolated to zero
power, is $\Gamma^{\rm theor}_{\rm tof}=550(5)$ Hz which
corresponds to the theoretical value for the averaged
time-of-flight broadening for the given delay $\tau=1210 \
\mu\rm{s}$ in the absence of other broadening contributions
(Fig~\ref{fig3}b).

The value $k_{\rm broad}$ can be tested in the same way as $k_{\rm
shift}$ with an uncertainty of  18\% (see Fig.~\ref{fig3}d). Due
to the complex structure of this value and the number of
contributions of different nature which add up to the total line
broadening, it is impossible to disentangle them and to set
separate restrictions on the transition matrix element and the
ionization cross-section. Nevertheless, if we switch off any of
the important broadening mechanisms, as illustrated in
Fig.~\ref{fig3}, the simulation will no longer agree so well with
the experimental observation. This means that both the
calculations performed in \cite{Haasetal} as well as the current
analysis adequately describes the excitation dynamics of the
hydrogen atom in the presence of photoionization.

For consistency of our analysis, we also performed an evaluation
of the laser line width $\Gamma_{\rm laser}$.  Studies of the
spectral laser line width at 486 nm, based on the investigation of
beat signals, resulted in a value of 60 Hz \cite{laserlinewidth}.
In the framework of this paper, we can make an estimate of its
spectral line width with a different, indirect method. To this
end, we compare the calculated line width at zero laser power
$\Gamma^{\rm theor}_{\rm tof}$ with the experimentally observed
line width $\Gamma_{\rm exp}$, where the latter is obtained by
averaging the extrapolated day values for the line width similar
to Fig.~\ref{fig2}b, resulting in $\Gamma_{\rm exp}=775(20)$ Hz.
 From Eq.~(\ref{LaserLinewidth})
we obtain $\Gamma_{\rm laser}=56(5)$ Hz, which is consistent with
the previous independent measurement \cite{laserlinewidth}. The
splitting of the $(1S,\, F=1, \,m_F=\pm1)\rightarrow
(2S,\,F'=1,\,m'_F=m_F)$ magnetic components is negligible in the
experiment.

An attempt to directly evaluate the expected Lyman-$\alpha$ count
rate in the given experimental geometry and thus test the
theoretical values more directly fails due to  the large
uncertainties in the beam density, the fraction of atomic hydrogen
in it and the detection efficiency.  The presented indirect analysis
happens to be much more sensitive to the absolute values of the
transition matrix elements and the ionization cross section than the
direct evaluations.

\subsection{Impact on the \sts\ Absolute Frequency Determination}

\label{chfreq}

 At the end of this section we  discuss the conclusions of
the current analysis regarding the absolute frequency measurement
of the \sts\ transition in atomic hydrogen.

First of all, we want to point out the non-linearity of the
frequency shift of the line center as a function of the laser
power (Fig.~\ref{fig3}a). Our analysis shows, that our previous
linear fit leads to an error of 10-20 Hz in the determination of
the absolute \sts\ frequency $\omega_{ge}/2\pi$, if we restrict
ourselves to the experimental range of powers ($0.1\; {\rm W}<P<
0.5\;{\rm{W}}$). This does not contradict the uncertainty of the
line shape model of 20 Hz  given in \cite{HuGrWeHa1999}. Moreover,
in the derivation of the possible small temporal drift of the
\sts\ frequency \cite{Fischetal} this error cancels out. Still,
for a further improvement of the \sts\ frequency measurement it is
highly desirable to reduce the  laser intensity and thus reduce
the error and the uncertainty of the extrapolation.

Secondly, the calculations show, that due to the photoionization,
the velocity distribution of atoms escaping the nozzle
significantly differs from the distribution of atoms contributing
to the signal. For the correction of the second order Doppler
effect it is more physical to model the initial velocity
distribution in the nozzle by a Maxwellian distribution (see
Eq.~(\ref{maxw})), which is dynamically modified by the ionization
process of atoms with long interaction times, instead of modelling
an effective velocity distribution at the detector. On the other
hand, the magnitude of the Doppler effect, which has been one of
the biggest concerns in the last measurements \cite{NiEtAl2000,
Fischetal} is already on the level of $-20$ Hz for a delay time of
$\Delta\tau=1210 \ \mu$s. This opens an opportunity to use a
simplified line shape model to correct for this effect.

 We have
performed an evaluation of the whole data set obtained in 2003
measurement \cite{Fischetal} using such a simplified approach. The
\sts\ spectral lines detected with the delays $\Delta\tau>1210\
\mu$s have been fitted by Lorentzian functions.  The frequencies
of the line centers have been corrected for the second-order
Doppler effect by adding the calculated frequency shift. The
result of this analysis was consistent with the result of the full
line shape model \cite{HuGrWeHa1999} within the uncertainty of the
model. The shift should be sensitive to the initial velocity
distribution which can differ from (\ref{maxw})  due to e.g.
geometrical factors, unperfect thermalization or collisions. The
delayed detection modifying the velocity distribution
(\ref{modmax}) drastically reduces the mean Doppler shift as well
as its sensitivity to the initial distribution. We tested
numerically, that the substitution of the $v^3$ dependency
(\ref{maxw}) by $v^4$  results in a mean Doppler shift of $-23$ Hz
for $\Delta \tau=1210\  \mu$s, which is only 3 Hz aside the
previous value of $-20$ Hz. The difference decreases for
increasing $\Delta\tau$ and the corresponding uncertainty in the
determination the \sts\ frequency can be reduced to a few hertz.
Thus, using such a simplified model we can correct the
second-order order Doppler effect with an uncertainty on the level
of a few parts in $10^{15}$.

\begin{figure} [t!]
\begin{minipage}[b]{0.49\textwidth}
\begin{center}
\includegraphics[width=0.85\textwidth]{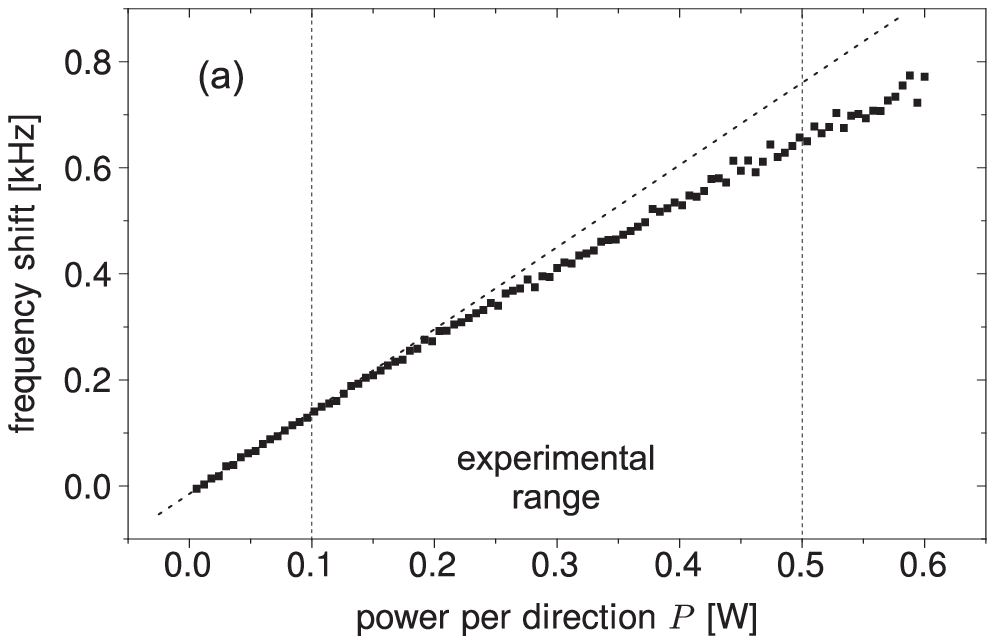}
\end{center}
\end{minipage}

\vspace{0.6cm}
\begin{minipage}[b]{0.49\textwidth}
\begin{center}
\includegraphics[width=0.85\textwidth]{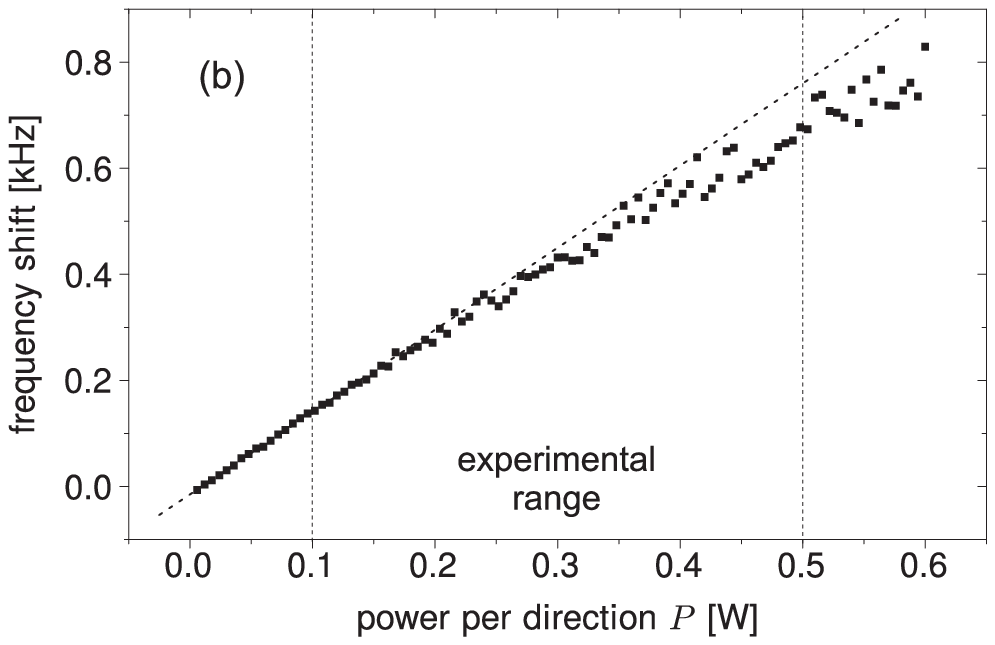}
\end{center}
\end{minipage}
\caption{Monte-Carlo simulation: (a) -- the frequency of the \sts\
transition vs. laser power for an ensemble of 10\,000 atoms for
the completely opened nozzle. The residual scatter of the data can
be explained by the stochastic nature of the Monte-Carlo
simulation. (b) -- the same simulation, but the nozzle radius is
randomly varied in the range \{$w_0$,\,$r_{\rm D1}$\}. We fit the
data at low intensity limit by linear functions (dashed tilted
lines) for visualization of non-linearity of the frequency shift.}
\label{figmartin}
\end{figure}

The last point concerns the influence of the freezing nozzle on
the observed absorption line centers, and resulting effects on the
absolute frequency, which is obtained by the extrapolation of
these line centers to zero laser power. Since the nozzle
temperature is lower than its melting point of 14 K, molecular
hydrogen effectively freezes on the nozzle walls and reduces its
diameter. In the experiment, we had to melt molecular hydrogen
approximately once per 20 min to avoid significant losses of laser
power and atomic hydrogen flux in the 243 nm cavity. Our
simulation allows for different spatial seedings of starting
points of the atomic trajectories at the diaphragms, which can be
used to model a reduced nozzle diameter or a misalignment of the
cavity mode with respect to the atomic beam axis.

 We have
performed a simulation of a whole day of measurement recording the
1210 $\mu$s delayed spectra at 100 different intensities, while
varying the nozzle radius randomly in the range from $r_{\rm D1}$
down to $w_0$. The resulting line center fits as a function of
laser power per direction are shown in Fig.~\ref{figmartin}b.
Figure~\ref{figmartin}a results from the same simulation but with
a permanently free nozzle. An uncontrolled freezing leads to an
excess scatter of the line centers, increasing for larger exciting
powers. Specifically, a smaller radius at the entrance diaphragm
results in an upwards shift in line center frequency, the shift
increasing with excitation power. This is consistent with the
observation in Fig.~\ref{fig3}c, where a small nozzle radius
results in a larger value for $k_{\rm shift}$. As a result, a
linear fit in the range of 0.1\,--\,0.5 W and an extrapolation to
zero intensity gives different values for the absolute frequency
depending on the amount of nozzle freezing. Typically the scatter
is on the level of a few tens of Hertz, which can only partly
explain the scatter observed in the experiments ~\cite{NiEtAl2000,
Fischetal}. Other contributions can originate in uncertainties of
the power axis and misalignment of the 243 nm enhancement cavity.
In future measurements we will aim at increasing the nozzle
diameter to avoid restriction of the atomic trajectories, working
at lower excitation powers, and improving the stability of the
enhancement cavity including power stabilization.

The beam nature of the \sts\ spectroscopy of atomic hydrogen
imposes certain restrictions on the experimental conditions like
high intensity of the excitation field, impossibility of further
cooling of the beam, short interaction time and freezing of the
nozzle. The enumerated effects result in a number of systematic
shifts and corresponding loss of accuracy. Note, that all optical
spectroscopic experiments demonstrating the accuracy better than
$10^{-14}$ deal with cold ions or atoms in traps (see e.g.
\cite{NIST, NPL}). All the systematic shifts of the \sts\
transition
 discussed in the paper would be significantly reduced or
eliminated if the hydrogen atoms were slowed down by laser cooling
or if one used hydrogen-like ions that could be held in a trap
\cite{Helium}.

Based on considerations of this Section we can conclude, that
perturbation of the mode of the enhancement cavity results in a
frequency shift via the intensity related shift and broadenings.
 Though the cavity is not confocal which
ensures the suppression factor over 500 for the TEM$_{01}$ mode,
the intra-cavity diaphragms can influence the structure of the
TEM$_{00}$ cavity mode. Monte-Carlo simulations performed for
impure TEM$_{00}$ mode shows that the results presented in the
article remain valid if the power falling into the TEM$_{01}$ mode
does not exceed 5\%. Thus, special attention should be paid for
intensity stabilization, coupling and alignment of the cavity to
avoid any additional non resonant light. After taking the issues
discussed in the paper into account it seems realistic to reduce
the uncertainty of the determination of the \sts\ frequency to the
level of a few parts in $10^{15}$.

\section{Conclusions}

In this article we discussed the transient excitation dynamics of
the \sts\ transition of atomic hydrogen under irradiation of two
counter-propagating 243 nm light beams. The main focus of the
present analysis has been the inclusion of photoionization of the
$2S$ state, power broadening and the inhomogeneous AC Stark in the
quantum dynamics and the detailed study of their influence on the
transition frequency and its line width.

Though  photoionization would only insignificantly change the line
width of each individual atom in the experimental range of
excitation powers and interaction times in a laser field of
constant intensity, the preferred ionization of slow $2S$ atoms
results in a substantial broadening of the observed collective
spectrum. The modelling shows, that the resulting
intensity-dependent frequency shift is mainly caused by the
dynamic Stark shift, while the intensity-dependent line broadening
results from power broadening, the dynamic Stark shift in a
spatially inhomogeneous laser field and the influence of
photoionization.

For our simulation we used the values of the transition matrix
elements, the dynamic Stark shift coefficients and the ionization
rates  re-derived in \cite{Haasetal}. The results of the
simulation show a fair agreement with the experimentally observed
ones. We can confirm the value of the difference of the dynamic
Stark shift coefficients $\beta_{\rm ac}(2S)-\beta_{\rm ac}(1S)$
with an accuracy of 15\% using the analysis of the \sts\ frequency
shift. The analysis of the line broadening also shows a good
agreement with the experimental values within 18\%. Due to a
complex structure of the intensity-dependent line broadening it
does not allow to make definite conclusions about the ionization
coefficient, though indicating that our treatment of the
excitation process in the thermal beam is correct.

The uncertainty of the determination of the \sts\ frequency is
determined not only by the experimental accuracy, but also by the
uncertainty of the line shape model and the correction of the
dynamic Stark shift. Taking photoionization into account allows to
build a more accurate line shape model. Factors such as
non-linearity of the line shift and geometrical factors contribute
on a level comparable in magnitude with the second order Doppler
effect, which has been of major concern in previous studies
\cite{HuGrWeHa1999, NiEtAl2000}. The Monte-Carlo approach
presented in this paper allows to perform accurate quantitative
evaluations of all these effects and make an accurate evaluation
of the final uncertainty.

\section{Acknowledgements}

N.K. acknowledges the support of  RFBR (grants No. 04-02-17443,
05-02-16801), the Alexander von Humboldt Foundation, and thanks
the scientific team of MPI f\"{u}r Kernphysik for the hospitality.
 U.D.J.~acknowledges support from Deutsche Forschungsgemeinschaft
(Heisenberg program).

\end{document}